# A geometrical framework for thinking about proteins

Jayanth R. Banavar[1], Achille Giacometti[2,3], Trinh X. Hoang[4], Amos Maritan[5], and Tatjana Škrbić[1,2]

[1] University of Oregon, Department of Physics and Institute for Fundamental Science, Eugene, Oregon, USA

[2] Ca' Foscari University of Venice, Department of Molecular Sciences and Nanosystems, Venice, Italy

[3] European Centre for Living Technology (ECLT), Ca' Bottacin, Dorsoduro 3911, Calle Crosera, 30123 Venice, Italy

[4] Vietnam Academy of Science and Technology, Institute of Physics, Hanoi, Vietnam

[5] University of Padua, Department of Physics and Astronomy, Padua, Italy

**Correspondence**

Jayanth Banavar
1258 University of Oregon, Eugene, OR 97403-1205, USA
Email: banavar@uoregon.edu

## Abstract

We present a model, based on symmetry and geometry, for proteins. Using elementary ideas from mathematics and physics, we derive the geometries of discrete helices and sheets. We postulate a compatible solvent-mediated emergent pairwise attraction that assembles these building blocks, while respecting their individual symmetries. Instead of seeking to mimic the complexity of proteins, we look for a simple abstraction of reality that yet captures the essence of proteins. We employ analytic calculations and detailed Monte Carlo simulations to explore some consequences of our theory. The predictions of our approach are in accord with experimental data. Our framework provides a rationalization for understanding the common characteristics of proteins. Our results show that the free energy landscape of a globular protein is pre-sculpted at the backbone level, sequences and functionalities evolve in the fixed backdrop of the folds determined by geometry and symmetry, and that protein structures are unique in being simultaneously characterized by stability, diversity, and sensitivity.

## Statement for broader audience

We present a simple geometrical model of a chain, which captures the essential features of globular proteins, and explore its consequences. Our model marries the ideas of Kepler, of objects touching each other, and Pauling, of hydrogen bonds providing scaffolding for helices and sheets. We suggest a poking potential for a chain, whose deployment yields the correct structures of both helices and sheets, while promoting the assembly of the building blocks into the tertiary structure.

## 1. Introduction

A globular protein is a designed heteropolymer, whose 'primary' sequence of amino acids, encoded by the DNA, is subject to evolution and is a molecular target of natural selection [1-4]. Proteins, the amazing molecular machines of life, are complex with myriad degrees of freedom. Linus Pauling [5,6] launched the field of molecular biology by developing the principles of quantum chemistry and applying them to predict the structures of protein modular building blocks, helices and strands assembled into sheets. Pauling and others, most notably Ramachandran [7] and Rose [8-12], adopted a backbone-based view, focusing on the role of the backbone atoms, the avoidance of steric clashes, and the importance of hydrogen bonds. A side chain centered view has highlighted the vital importance of the role of the solvent, the distinct hydrophobicity/hydrophilicity of amino acid chains, and the need to sequester the hydrophobic core from the solvent, resulting in an elegant picture of a folding funnel landscape [13-15]. Our work here is built on efforts over the last two decades, using ideas from geometry and symmetry [16-37] that view a protein as a tube of non-zero thickness.

Proteins are distinctive chains with many special attributes. They are complex – there are twenty types of amino acids with side chains with distinct physical and chemical attributes. The behavior of a protein is governed by myriads of interactions amongst the constituent atoms and the surrounding water molecules. These interactions include van der Waals forces, hydrogen bonding, electrostatics, hydrophobicity mediated by the surrounding water molecules, and the imperative need to avoid steric clashes. Despite this bewildering complexity, globular proteins share an impressive array of common characteristics. Small globular proteins fold rapidly and reproducibly into their native state structures [38]. The native state folds are evolutionarily conserved [39-41] and are immutable. All protein native state structures are made of common building blocks: helices and zig-zag strands assembled into almost planar sheets. Many protein sequences adopt the same native state conformation. The native state structure of a protein is robust to significant amino acid mutations [42,43] except at certain key locations [44-46]. Under some circumstances, protein chains exhibit a tendency to aggregate creating water insoluble amyloid [47-49]. Such amyloid formation is implicated in debilitating diseases. One cannot but wonder whether these common attributes of globular proteins reflect a deeper underlying unity in their behavior.

A great simplification in determining protein structure is the neat separation between the roles of backbone and side chain atoms. Hydrogen bonding between the backbone atoms is largely responsible for the creation of the common 'secondary' building blocks of protein structure. The side chains stay out of the way in both building blocks but play an important role during their assembly into the folded 'tertiary' state. Protein structures are modular, and their assembly is facilitated by loops, which are non-repetitive structural elements [50-55]. Because of the modularity of protein structures, the total number of distinct native state folds total just several thousand [56-59] in all instead of the vastly larger number that one would expect for a featureless conventional chain molecule of this length. The geometries of the native state structures provide the context for the variety of interactions between proteins and other cell products. In our picture, the textbook wisdom [4] that 'sequence determines structure' is changed to 'a sequence chooses its native state structure from a menu of modular native state structures comprised of the common building blocks of all globular proteins' [20]. The number of distinct folds is significantly smaller than the number of protein-like sequences that fit into them just as the number of items on a restaurant menu is typically much smaller than the number of patrons, providing an explanation for why machine learning is wonderfully suited to, and enormously successful in, matching a sequence to its native state structure [60-62]. Indeed, recent work has demonstrated the successful determination of atomic structure from a single sequence with a large protein transformer language model, with 15 billion parameters, without the need for evolutionary information present in multiple sequence alignments [63].

We will build on the premise that the determination of the native state structure of a protein sequence is a two-step process [20]. All proteins share a common backbone, and, in the first step, the interactions between backbone atoms create the building blocks of protein structures independent of the amino acid sequence. In the second step, the specific side chain interactions choose the best fit assembled native state structure from the menu of topologically distinct folds, already presculpted at the backbone level.

As noted earlier, hydrogen bonding plays a major role in secondary structure formation. However, it is no longer the sole or even the dominant interaction promoting the assembly of the tertiary structure. One may consider an isotropic attractive interaction, that aims to surround a $C_\alpha$ atom with as many others as possible within a given range, as a surrogate for the plethora of actual interactions [64,65]. However, this would conflict with the specific anisotropic action of the hydrogen bonds. Indeed, common sense suggests that any kind of generic isotropic attraction, mimicking the hydrophobicity mediated by the water, would destabilize both the topologically one-dimensional helix and strand into three-dimensional compact structures. The challenge is to determine how the constraints imposed by the common backbone attributes yielding the pre-sculpted landscape along with sequence specificity compatibly yield the choice of the most appropriate fold. Here we identify a simple way of capturing the emergent interactions in both steps of the two-step process in a harmonious manner. Our overarching goal is to elucidate the simplest set of governing principles that dictate protein behavior.

## 2. Materials and Methods

### 2.1 Protein geometry

Our approach is informed by some gross features of empirical data on proteins acquired over the decades and stored in the PDB [66]. These features are not as detailed as those used by Pauling [5,6] in his pioneering work, and we will state what these are as we go along. Our model is specialized for a protein, even though some of the ideas we introduce here may find applications elsewhere. Unlike a protein with myriad interactions, a virtue of our emergent model is its sheer simplicity. Our goal is to capture the common characteristics of globular proteins in a tractable coarse-grained model whose assumptions are clearly stated and whose consequences can be deduced straightforwardly. We validate the model by exploring its predictions and benchmarking them with data.

Our model of a chain is inspired by past path-breaking studies. Hard spheres, each with the same radius, are the simplest emergent entities for modeling matter. The centers of a pair of spheres that just touch each other are separated by a distance equal to the sphere diameter. In the thermodynamic limit (when the number of spheres is infinitely large), one arrangement, which maximizes the number of pairs that just touch, is a face-centered-cubic (fcc) crystal [67-69] with each sphere touching 12 others. An optimal arrangement at one location will be optimal at any other and this leads to periodicity and translational invariance, signatures of a crystalline phase. Maximizing the packing fraction can be thought of in this context as being equivalent to a space-filling configuration.

In the protein arena, Corey, Pauling, and later Koltun, pioneered the use of a calotte space-filling model, now called the CPK model [70,71], in which the atoms are represented by spheres. The notion of space-filling was also explored in detail by Richards [72-74], who highlighted the relationship between a space-filling native state structure and the desire to occlude any hydrophobic surface from the surrounding water.

Figure 1a shows the packing of the protein myoglobin as viewed in a CPK model. The beauty and simplicity of protein structure is underscored in Figure 1b showing, that lurking underneath this complexity, the chain of backbone $C_\alpha$ atoms wends its way through structured helices linked by loops. Figures 1c and 1d show the same protein structure in a tube representation. The tube has been drawn with a diameter of 5.26Å corresponding to the prediction of our theory that we will present later in the paper. All but a few heavy backbone atoms are completely enclosed within the tube (Figure 1c). The atoms of the larger side chain atoms sticking out of the tube are seen in Figure 1d. Many of the atoms of the smaller side chains are completely enclosed within the tube. The situation in proteins is more complex compared to Kepler's packing of cannonballs [67-69] or a grocer's packing of apples because of the distinct sizes of the atoms and the tethering between them. As noted by Richards [73], *"for chemically bonded atoms the distribution is not spherically symmetric nor are the properties of such atoms isotropic"*.

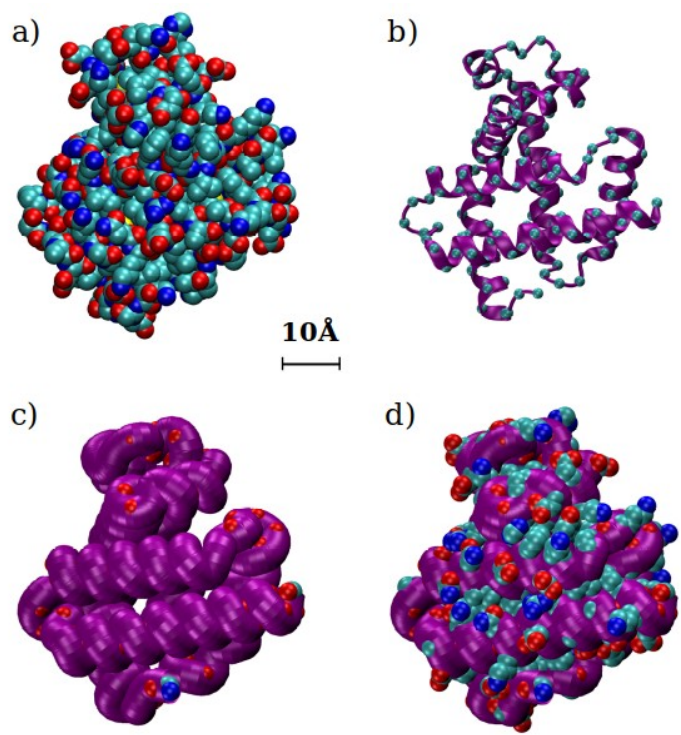

Figure 1: Panel a) Native state of myoglobin (PDB code: 3RGK) in the CPK representation in which all heavy atoms of the protein backbone and its side chains are represented as spheres with radii proportional to their respective van der Waals atomic radii. Color code: carbon (cyan), oxygen (red), nitrogen (blue), and sulfur (yellow). Panel b) Peering beneath the complexity. The myoglobin native state structure is shown in ribbon representation (in purple) with cyan spheres, shrunk in size for the sake of clarity, at the positions of the $C_\alpha$ atoms. Panel c) The myoglobin structure shown in a tube representation (also in purple) with the tube diameter chosen to be the theoretically predicted value of 5.26Å. The backbone oxygen atoms (red spheres) not entirely enclosed by the tube are visible. Panel d) The same tube representation but this time depicting the backbone and side chain atoms not fully enclosed by the tube.

## 2.2 Curation and data analysis

Our data consists of 4391 PDB structures, a subset of Richardsons' Top 8000 set [75] of high-resolution, quality-filtered protein chains (resolution < 2Å, 70% PDB homology level), that we further filtered to exclude structures with missing backbone atoms, as well as amyloid-like structures. For the analysis of protein helices presented here, we identified 3594 non-overlapping segments 12-residues long with coherently placed backbone hydrogen bonds between residue i and both i-4 and i+4 within the segment. For the analysis of protein β-sheets we identified 8422 antiparallel pairs of strands, by detecting three inter-pair hydrogen bonds at (i,j), (i+2,j-2), and (i-2,j+2), with i and j belonging to the two strands respectively; 4542 parallel strand pairs were identified by four inter-pair hydrogen bonds between (i,j-1), (i,j+1), (i+2,j+1), and (i-2,j-1). Double counting of the pairs was carefully avoided. Hydrogen bonds were identified using DSSP [76].

## 2.3 Details of computer simulations

We have employed two distinct Monte-Carlo methods in our simulations to obtain ground state conformations of our model: microcanonical Wang-Landau (WL) simulations [77] and replica exchange (RE) (or parallel tempering) canonical simulations [78]. We used both methods to check for consistency. The WL method is based on the iterative filling of energy histograms thus allowing us to estimate the density of states of the system. The acceptance probability in the WL ground state search is chosen to promote moves exploring less populated energy states and seeking to flatten energy histograms over the course of the runs. The RE approach consists of canonical simulations in parallel over a wide range of temperatures that bracket the 'transition temperature' between the folded and unfolded states, while concentrating in the low temperature region to search for low-lying states. Each simulation provides a replica of the system in thermal equilibrium. The swapping of replicas allows for rapid search. Both methods employed standard local moves including crankshaft, reptation, and endpoint moves along with the non-local pivot move.

## 3 Results and Discussion

### 3.1 Holding hands in a helix

We model the backbone of a protein as a chain of $C_\alpha$ atoms with a constant bond length b of 3.81Å [79]. Our initial focus will be just on the backbone atoms and their role in the sculpting of the building blocks. We identify the $C_\alpha$ atoms with the labels 1, 2, 3… i-1, i, i+1…. Each internal site i has two special directions, one from i to i-1 and the other from i to i+1. The local direction of the chain breaks the spherical symmetry and leaves, in the simplest scenario, a residual cylindrical symmetry that is preserved by a coin rather than a sphere. Therefore, the *simplest* geometry of objects with the correct symmetry at site i is a pair of intersecting coins of uniform radius Δ and infinitesimal thickness, both centered at i. The uniform radius assumption is justified because all proteins share the same backbone independent of the side chain specificity. The normal directions of the two coins at site i are chosen to indicate the directions to the nearest neighbor sites of i.

Following Pauling [5,6], we wind our chain into a helix. We treat the residues as being equivalent, ignoring the differences in the side chains, which are not our current focus for understanding secondary structure formation. To go from one residue to the next, one would rotate about the helical axis by an angle $\varepsilon_0$, while simultaneously translating along the axis by the rise per residue p (see Figure 2 for a sketch). For a fixed bond length b, a helix is completely characterized by $\varepsilon_0$ and p. The translation along the axis for a complete turn of 360° is the pitch of the helix P. A generic helix is three dimensional and is a curve with a *constant* radius of curvature [80]. However, for $\varepsilon_0 = 180°$, the helix becomes a zig-zag two-dimensional strand. Defining the helix radius to be R, the Cartesian coordinates of the points lying along a helix are, with integer n,

$$x_n = R \cos((n-1)\varepsilon_0)$$
$$y_n = R \sin((n-1)\varepsilon_0)$$
$$z_n = P(n-1)\varepsilon_0/(2\pi)$$

The radius of curvature of a continuous helix is given by:

$$R_{curv} = R(1+\eta^2),$$

where $\eta = P/(2\pi R)$.

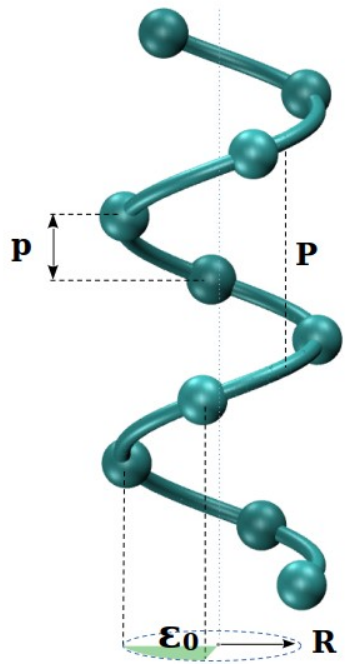

Figure 2: Sketch of a discretized helix with uniform bond length between successive points. The figure shows the rotation angle per bead $\varepsilon_0$, the rise per bead p, the helix pitch P and the helix radius R.

Pauling and his collaborators [5] determined a hydrogen-bonded helical configuration for the polypeptide chain. Here we follow in Pauling's footsteps, but inspired by Kepler, we now adopt a geometrical approach to find an optimal helix (we denote this as a Kepler helix) in which every coin in the interior of the helix just touches another backbone coin with the centers of the partner coins separated by a distance exactly equal to the coin diameter 2Δ (Figure 3).

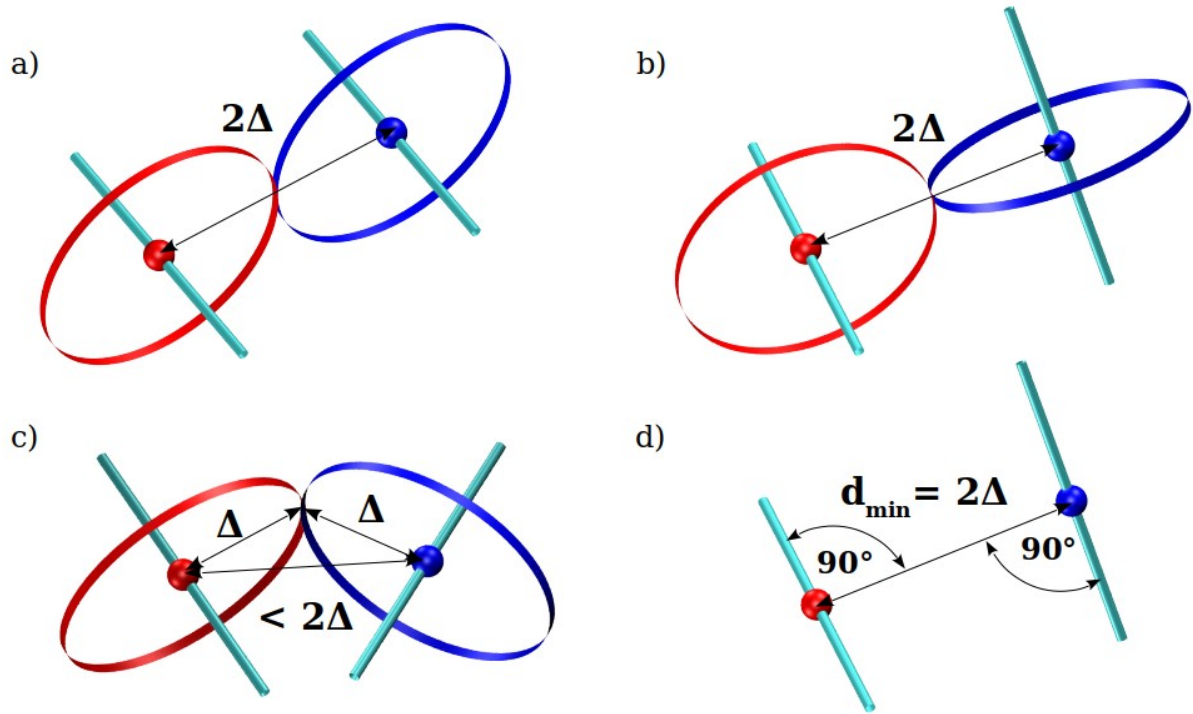

Figure 3: Sketches of touching coins and the constraint on the distance between coin centers. Figures 3a and 3b show the canonical touching of a pair of coins. The main difference is that in the first panel the two coins lie in a plane whereas in the second they do not. In both cases, the distance between the coin centers equals the coin diameter. Figure 3c depicts an example of two coins just touching but with the distance between their centers less than the coin diameter. We do not consider such conformations here as legitimate touching. Figure 3d is an illustration of the geometrical criterion for determining the closest distance between two skew lines (the ones depicted in Figure 3b) highlighting the need for both distance and angle constraints. In practice, for Kepler handholding in a discrete helix, the (i,i+3) distance ought to be equal to the coin diameter and the (i,i+3) straight line needs to be perpendicular to both the (i-1,i) and (i+3,i+4) straight lines. These types of Frenet constraints were studied earlier by Hoang et al. [20] and form the basis of computer simulations in that work and here.

Figure 4 is a sketch of the *unique* Kepler helix for the case in which the two coins at site i just touch partner coins closest in sequence along the chain, one at i-3 and the other at i+3, for every i in the helix interior. More specifically, the coin at (i-3) with its direction pointing towards (i-4) just touches the coin at i with its direction pointing towards (i+1). Likewise, translating by 3, the coin at i with its direction pointing towards (i-1) just touches the coin at (i+3) with its direction pointing towards (i+4). There is a steric constraint on coins along the helix that the centers of pairs of coins at distinct sites (that are not nearest neighbors) are spaced farther apart than the (i,i+3) distance.

In the continuum limit, the Kepler helix becomes the *space-filling* helical conformation of a tube [16]. There is then an equality of the radius of curvature of the helix and the minimum non-local three body radius (a measure of the closest approach of two parts of a tube) signifying space-filling. The two coins at site i now overlap completely. A tube may be viewed as a chain of coins in the continuum limit (the coins maintain their radius Δ but get closer and closer to each other) with Δ= $R_{curv}$. To get a space-filling continuum helix, one would take a tube and bend it as tightly as possible locally while avoiding any kink and place successive turns on top of and alongside each other. There are no intersections and when viewed from the top, there is no hole in the middle. Nor is there any space between successive turns. In the continuum limit, the bond length tends to zero, the bond bending angle, θ, tends to 180°, the rise per residue, p, approaches 0, and the rotation angle, $ε_0$, tends to 0. One can only deduce the key geometrical dimensionless pitch to radius ratio in the continuum case, but not other relevant quantities like the coin radius, because there is no characteristic non-trivial length scale like the bond length.

We will choose Δ=$R_{curv}$ leaving us with just one characteristic length scale in the Kepler helix. Simple geometrical considerations[1] dictate that, for every i, the (i-3,i) distance is equal to the coin diameter 2Δ and the [(i-3),i,(i+1)] and [(i-4),(i-3),i] angles both equal 90° (see Figure 3). The centers of coins at non-contiguous points are farther apart than 2Δ and cannot intersect. This is what we refer to as handholding in a helix with each $C_\alpha$ atom having two hands (coins) and every hand in the helix interior holding another hand. The sidechains stick out in the approximately negative normal direction and stay out of the way of the backbone atoms and each other in the Kepler helix.

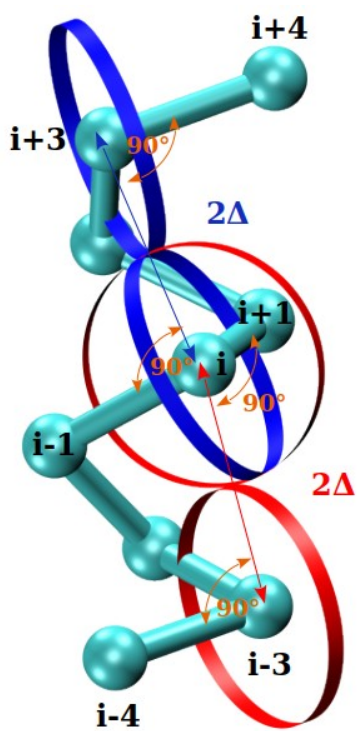

Figure 4: Sketch of the Kepler helix (see Supplementary video). The two coins (in red and in blue) at bead i are shown. One coin is shown for bead i-3 (in red) and for i+3 (in blue). These touch the two coins at bead i. The pair of blue coins touch each other as do the pair of red coins. The distances of 2Δ (the coin diameter) and the angles of 90° characterizing the geometrical conditions of touching are indicated. Every pair of non-contiguous coins that do not touch is farther than 2Δ from each other and therefore does not intersect.

We note the accord between the parameters characterizing the Kepler helix, the Pauling hydrogen-bonded helix, and protein helices (see the upper half of Table 1). The Kepler helix is *not* space-filling (one would need an infinite number of two-dimensional coins to fill three-dimensional space). The local bond bending angle of the Kepler helix, $\theta$, is found to be around 91.8° suggesting that the minimum sterically allowed bending angle, $\theta_{min}$, ought to be smaller than but close to that value. Of course, the spread in the geometries of the 20 amino acids, especially the presence of small amino acids like glycine, ought to allow for tighter bending, facilitating turns in the protein structure.

We can estimate from the steric requirement, the (i,i+2) distance must be greater than $2\Delta$, that $\theta_{min} = 2 \sin^{-1}(\Delta/(2b)) \sim 87.3°$. The coin radius $\Delta$ is deduced to be around 2.63Å for the Kepler helix from our calculations. Remarkably, Pauling's hydrogen bond analysis and the purely geometrical deductions yield consistent results for the helix geometry. As a bonus, we can determine the coin (tube) radius $\Delta$ and the tightest bond bending angle $\theta_{min}$, with no additional assumptions, from the properties of the Kepler helix. Figure 1 illustrates that the theoretically predicted value of $\Delta \sim 2.63$Å provides enough space to hold the backbone atoms and atoms of the smaller sidechains within it.

| | Kepler helix | Pauling helix | Protein helices |
|---|---|---|---|
| Rotational angle $\varepsilon_0$ [°] | 99.8 | 97.3 | 99.1 ± 3.4 |
| Rise per residue p [Å] | 1.58 | 1.47 | 1.51 ± 0.08 |
| $2R_{curv}$ [Å] | $2\Delta = 5.26$ | 5.32 | 5.25 ± 0.28 |
| | Kepler anti-parallel strands | | Antiparallel β-sheets in proteins |
| d(i,j)[Å] | $2\Delta = 5.26$ | | 5.26 ± 0.20 |
| $d(M_i,M_j)$[Å] | $< 2\Delta$ | | 4.31 ± 0.22 |
| | Kepler parallel strands | | Parallel β-sheets in proteins |
| $d(i,M_j)$ [Å] | $2\Delta = 5.26$ | | 5.26 ± 0.16 |
| $d(M_i,j)$ [Å] | $< 2\Delta$ | | 4.90 ± 0.31 |

The bond length b=3.81Å. Our theory predicts Δ of around 2.63Å.

Table 1: (Upper half) Comparison of the geometries of the Kepler helix, the Pauling helix (α-helix with 3.7 residues per turn), and protein helices. All attributes of a helix can be deduced from the bond length b, the rotation angle $\varepsilon_0$, and the rise per residue p. The table also shows the value of twice the radius of curvature of the helix, $2R_{curv}=2R(1+\eta^2)$, which is a measure of coin diameter in our theory. The geometries of the Kepler helix and the Pauling helix are compatible with each other and with empirical data, within the error bars. The Pauling helix is derived using input of quantum chemistry unlike the Kepler helix.

(Bottom half) Comparison of the geometries of the two types of arrangements of Kepler strands and the geometries of parallel and antiparallel β-sheets in proteins. In the case of the anti-parallel Kepler strands, the pair of touching coins (whose centers are beads i and j, see Figure 5b) are of the same color, while in the case of the Kepler parallel strands, the pair of touching coins (having the centers at points i and $M_j$, see Figure 5c) are of different colors. $M_j$ is defined to be the geometrical center of beads at positions j-1 and j+1 (see Figures 5b and 5c). Note that unlike the uniaxial helix, a zigzag strand is biaxial (see text below and Figures 5b and 5c). This flexibility permits an effective squeezing of strands in a sheet thereby promoting more compact packing while yet maintaining the touching conditions between the coupled axes. This squeezing is reflected in a smaller mean distance $d(M_i,M_j)$ compared to the mean distance $d(i,j)$ in antiparallel sheets (i and j belong to the coupled axes in the antiparallel case), as well as in a smaller mean distance $d(M_i,j)$ than the mean distance $d(i,M_j)$ in parallel sheets (here i and $M_j$ belong to the coupled axes).

Protein helices are predominantly right-handed because the amino acids themselves are left-handed. This chiral symmetry breaking originates from steric clashes of oxygen backbone atoms with the side chain atoms in a left-handed helix [7,8]. We note that, just like in the Pauling analysis, there is no spontaneous chiral symmetry breaking in our model. Operationally, one can break the symmetry by hand as we will, in our simulations, to exclude extended left-handed helices.

It is useful to recapitulate what we have done. We started with a discrete chain of uniform bond length 3.81Å curled into a helix and did not consider the role of side chains. We assigned two coins, each having a radius equal to the helix radius of curvature, to each interior site with their normal vectors pointing towards its neighbors. We then determined both the geometry of the optimal Kepler helix and thence the coin radius by requiring that every coin in the helix interior touched a partner coin three apart along the sequence and that (i,i+3) was the closest non-contiguous pair along the helix. There were no other assumptions or adjustable parameters. The Kepler helix is in good accord with the Pauling helix and protein helices. There is no reason why this should necessarily be the case, and this may very well be coincidence. But we will take this accord seriously and explore other consequences here by building on the Kepler helix idea.

The analysis of the optimal helix teaches us several lessons. Each backbone $C_\alpha$ atom is endowed with two hands corresponding to the two coins. This sets a limit on the maximum number of backbone-backbone interactions of the coin-touching type at 2 per each $C_\alpha$ atom. Intriguingly, hydrogen bonding at the backbone level originates from two hands as well: a donor (-NH amide group) and an acceptor (-C=O carbonyl group) in the peptide main chain.

It has been pointed out by Rose [8] that "*all globular proteins (with the occasional exception of small, metal-binding polypeptides or those stabilized by disulfide bridges) are built on backbone scaffolds of α-helices and/or strands of β-sheet, the only two conformers where, with minor exceptions, the number of donors and acceptors is exactly balanced.*" Armed with this insight, we now turn to an analysis of touching coins within the second building block of a sheet, transferring knowledge from the Kepler helix of the coin radius Δ = 2.63Å.

## 3.2 Holding hands in a sheet

A generic helix is three dimensional. However, as noted earlier, when the rotation angle $\varepsilon_0$ = 180°, the helix becomes a zig-zag two-dimensional strand (Figures 5 and 6). In a helix, the handholding was local, sequentially separated by 3. The resulting Kepler helix is rigidly constrained. There ought to be more latitude within a sheet. First, unlike a tightly curled helix, the bond bending angle θ no longer needs to be as tight. In fact, θ tends to be large and exhibits considerable variation [79]. Similarly, there is some small variability around 180° in the dihedral angle μ [79] ensuring local planarity yet allowing for the strand to twist. The price paid for this flexibility is that a strand needs partner strands to hold hands with and this is necessarily non-local. But, unlike in a helix, one might hope that there can be coordinated hand holding in a sheet (Figures 5 and 6). Figure 5a shows a pair of identical ideal strands a distance 2Δ apart. The directions of the pair of coins at sites i and j are towards their neighbors, exactly as in the helix. Each of the two coins at site i touch the two coins at site j. The structure is an idealized version of a β-hairpin in which the two strands run antiparallel to each other.

Nature is clever and she throws us a curveball by coming up with variations to accommodate parallel strands in addition, provide flexibility, and enable squeezing and compaction. A strand within an idealized sheet is strictly two-dimensional and can be divided into two sub-lattices made up of every other $C_\alpha$ atom (Figures 5b, 5c and 6). Each sublattice forms a straight line which can be thought of as the axis of a straight tube or cylinder.

Nature adapts our assumptions pertaining to coins in a helix to the strand case, where she now works with every other site and straight axes. Because of the straight-line geometry of an axis, the two coins at a given site merge into a single coin – the (i,i-2) direction coincides with the (i,i+2) direction in an ideal strand. Noting that there are two distinct axes associated with a strand and two identical hands at the same location is superfluous, Nature effectively retains one coin at site i and moves the other coin with the same orientation to site $M_i$ (defined as the mid-point of i-1 and i+1, which lies on the other axis not passing through i) (Figures 5b and 5c). In this way, the site i can now act as a representative of either axis of the strand either through itself or its virtual image $M_i$. The coin touching condition is exactly as before with the two pairing axes (one from each partner strand) being parallel and 2Δ apart (Figures 5b, 5c and 6a).

A strand utilizes one of its axes for pairing with a partner strand with the other axis available for a second handholding partner axis or being free and unencumbered. Such flexibility permits an effective squeezing of strands in a sheet thereby promoting more compact packing while yet maintaining the touching conditions between the coupled axes (Figure 6d). This is in fact what is observed in proteins (see bottom half of Table 1). More fundamentally, the biaxial sheet has greater flexibility than in a helix allowing for the structure to cleverly adjust the lengths of hydrogen bonds, the degree of squeezing, and the dihedral angles determining the local twists in a strand [81-84] to accommodate favorable interactions, including those between sidechain atoms.

We predict that the displacement between adjacent zig-zag strands, tracking and in phase with each other, can take on a value of either 2Δ (for antiparallel strands, see Figures 5b and 6a) or a smaller value depending on the strand bond bending angle (for parallel strands, see Figures 5c and 6b and the bottom half of Table 1). In the antiparallel case, as in a hairpin, the donor on one strand can adjust to face an acceptor on another allowing for horizontal ladder like hydrogen bonds. In contrast, for the parallel case, a donor is necessarily across another donor and thus one must have zigzag hydrogen bonds allowing for a strand separation of less than 2Δ to accommodate the same hydrogen bond length. For the antiparallel case, a coin at i touches one at site j directly across it, whereas for the parallel case, the coin at i touches the coin at $M_j$ (Figures 5b and 5c). In the two cases, the (i,j) and the (i, $M_j$) distances are predicted to be 2Δ or 5.26Å respectively (see the bottom half of Table 1). As in the helix, the sidechains stay out of the way in the sheets and do not play a role in the consideration of the backbone coins holding hands.

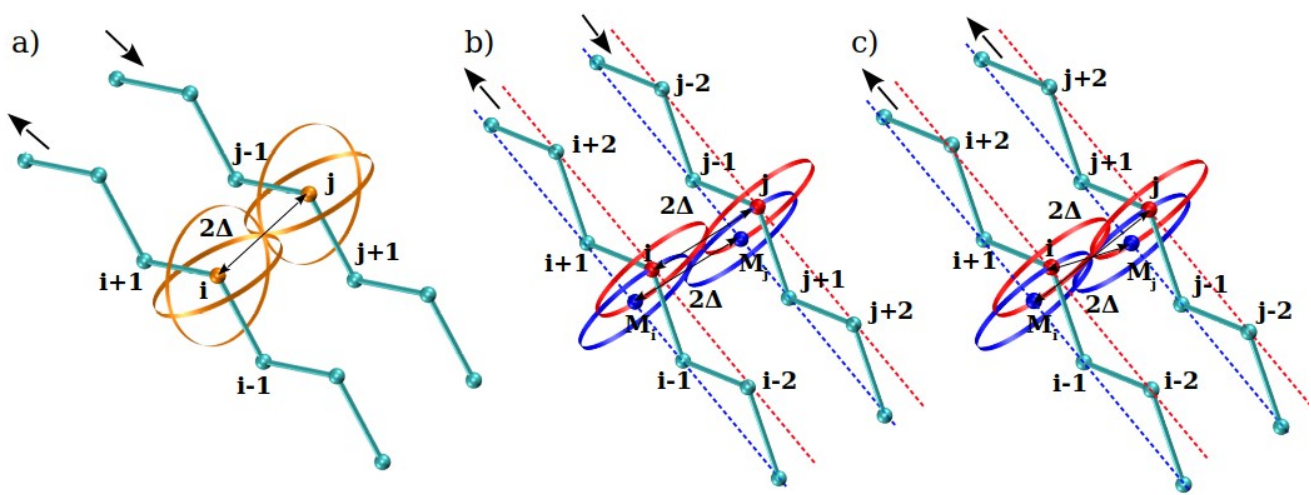

Figure 5: Distinct possibilities for the coordinated handholding for a pair of identical ideal strands. Panel a) shows a pair of strands a distance 2Δ apart with associated pairs of coins (shown in orange color) at sites i and j that are oriented towards their neighbors along the chain (i-1 and i+1, for bead i, and j-1 and j+1, for bead j), exactly as in the Kepler helix. Each of the two coins at site i touch the two coins at site j. The (i-j) distance is 2Δ. Panel b) shows the same structure as in a). Unlike the uniaxial helix, a zigzag strand is biaxial. Each strand has two axes. The blue axis of the left strand goes through sites (i-1, $M_i$, i+1) whereas the red axis passes through the points (i-2, i, i+2). Here $M_i$ is the mid-point (i-1,i+1). The red coins of i and j touch each other as do the blue coins at $M_i$ and $M_j$. The (i-j) distance is again 2Δ. Panel c) shows a distinct conformation of the two strands. The red coin at i just touches the blue coin at $M_j$. The (i-$M_j$) distance is now 2Δ and the (i,j) distance is smaller. Conformations b) and c) correspond to idealized antiparallel and parallel arrangements of strands.

We have rationalized the formation of the common building blocks of all proteins by imposing specific geometrical constraints involving the backbone atoms of a protein. We have sought to maximize systematic handholding of pairs of uniform size coins of radius Δ with well-defined orientations (for the helical case) or the pairs of strand axes (for the sheet case). We will show in a companion paper that the theoretical underpinnings as well as the geometries of the protein building blocks are in good accord with protein data. Our results suggest that not only hydrogen bonds but also holding hands can independently provide scaffolding of both helices and sheets. The fundamental lesson is that, in the building blocks, holding hands heightens happiness, as in real life.

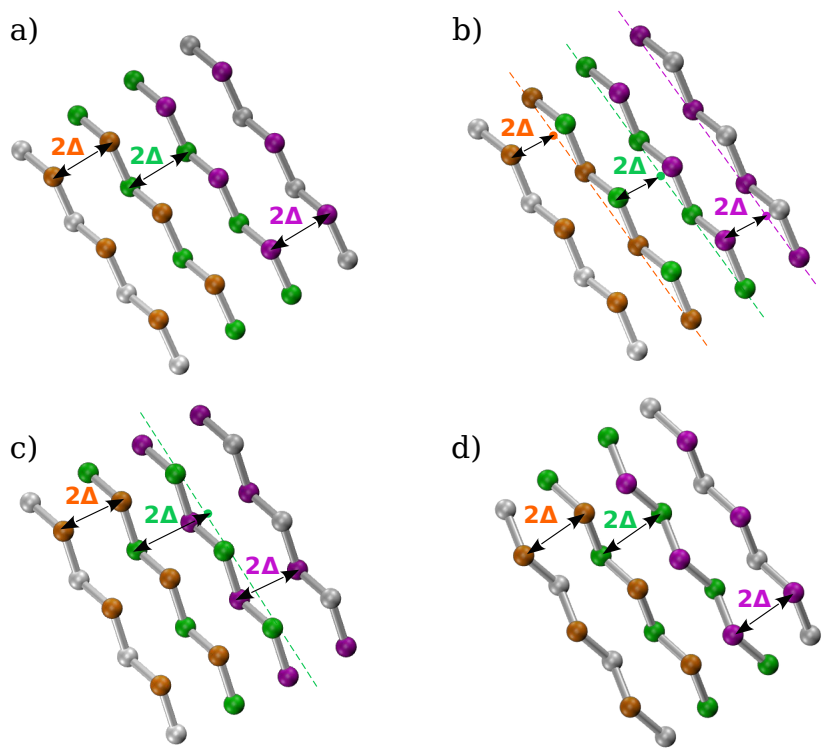

Figure 6: Four geometries of a sheet comprising idealized strands. In all cases, the axes coupled are orange-orange, green-green, and purple-purple with a spatial separation of 2Δ. Panel a): Antiparallel chains have spacings between the strand pairs equal to 2Δ. Panel b): Parallel chains have spacings between the strand pairs now closer than 2Δ. This is because the pairing axes go through the 'mountains' on one strand and 'valleys' on the other. Panel c) Mixed arrangement of four ideal strands. Antiparallel strand spacings are again 2Δ but the spacing between parallel chain segments 2 and 3 is less. Panel d): Antiparallel arrangement of four strands that depicts the squeezing of the sheet promoting its compaction while yet respecting the touching conditions. The red point in the second strand from the left is now closer to the corresponding purple point in strand 3 from the left than 2Δ unlike in Panel a). Interestingly, the hydrogen bonding patterns are ladder like for antiparallel chain segments and zigzag for parallel chain segments accounting for the distinct distances between axes.

## 3.3 Poking contacts and nestling promote assembly of tertiary structure

Unlike unconstrained objects, the location of an object must be supplemented with information about its context within the chain. If two objects along a chain happen to be close to each other, that does not necessarily mean that they have an affinity to each other – proximity does not equate to affinity. This is readily obvious for the case when two particles happen to be tethered next to each other along the chain. Their proximity does not imply anything about their liking each other or not. Likewise, if two particles from different parts of a chain are close by, that could very well be because the true affinity is between one of the particles and a neighbor along the chain of the other particle.

The backbone coin interactions of the backbone atoms were simple to deal with because we used coins all the same size with known orientations. Following Kepler and Pauling, we were able to work out the geometries of backbone conformations that allowed for the systematic touching of coins. The situation is murkier for interactions mediated by sidechains. This is because side chains have a range of geometries and chemistries and there is not any simply defined, let alone universal, object or orientation describing all of them. Furthermore, the plethora of interactions at the sidechain level makes the situation truly complex. The outcome of this complexity is nevertheless a simpler physical picture of compatible and complementary sidechains nestling together availing of their mutual attraction while aiming to exclude water from the hydrophobic core.

To capture this complexity in a simple, albeit approximate, manner, we will glibly continue to *ignore* side chains and introduce the concept of *poking pairwise interactions* between $C_\alpha$ atoms i and j, located at **r**$_i$ and **r**$_j$ respectively, satisfying the distance, d(i,j), criteria:

$$d(i,j) < d(i,j-1)$$
$$d(i,j) < d(i,j+1)$$
$$d(i,j) < d(i-1,j)$$
$$d(i,j) < d(i+1,j)$$

(Figure 7). These poking contacts identify significant pairwise interactions in a designed heteropolymer like a protein and indicate true affinity between i and j. As seen in Figure 7, i and j protrude towards each other and are prime candidates for touching in comparison to the 4 other nearby pairs (i,j-1), (i,j+1), (i+1,j) and (i-1,j). Indeed, in the Kepler helix, every (i,i+3) contact is a poking contact. Also, in an idealized sheet comprising in-phase strands alongside and tracking each other, there are poking relationships between every pair of touching coins in both parallel and antiparallel pairing. Indeed, common sense dictates that two parts of chain that are strongly attracted to each other must poke towards each other. This is a necessary corollary for a chain topology and is not relevant or even defined for unconstrained particles.

We have analyzed 3594 α-helices, 8422 antiparallel pairs of β-strands, and 4542 parallel pairs of β-strands in 4391 proteins of our data set (see Section 2.2) and have found that the presence of scaffolding hydrogen bonds in the interior of the common building blocks of proteins are associated with poking interactions in ~97% of cases in α-helices, ~95% of cases in antiparallel pairs of β-strands, and in ~99.9% of cases in parallel pairs of β-strands.

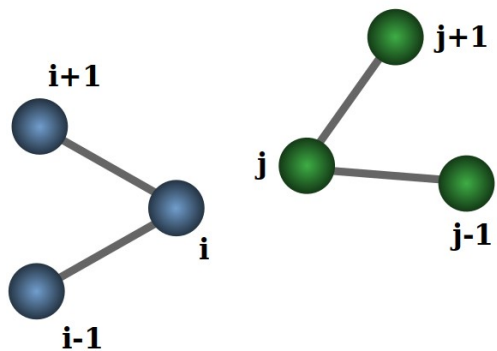

Figure 7: Illustration of two snippets of a chain depicting a poking pairwise contact between i and j. i is closer to j than to the two neighbors of j and likewise for j.

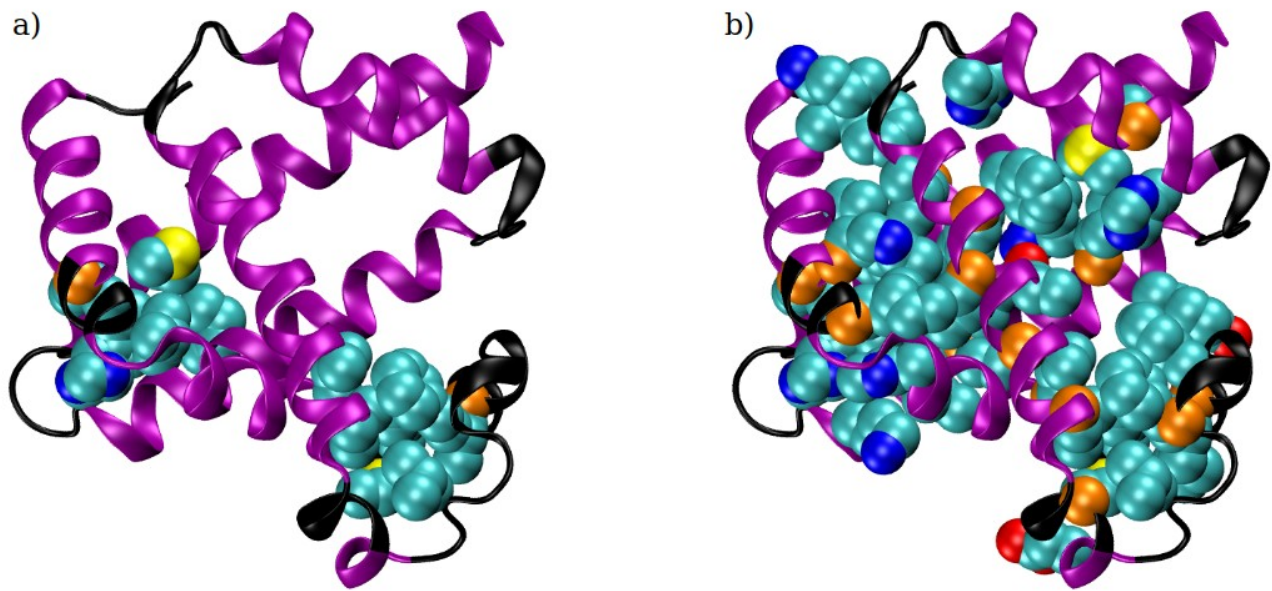

Figure 8: Nestling of side chains in the native state of myoglobin (PDB code: 3RGK). Both panels show the seven myoglobin helices (in purple) and its loops (in black). The atoms are drawn in CPK representation, in which the radii of the spheres correspond to the van der Waals radii of the respective atom types. Color code: backbone carbon atoms (orange); sidechain atoms: carbon (cyan), nitrogen (blue), oxygen (red), and sulfur (yellow). We begin by identifying the poking pairwise contacts between backbone $C_\alpha$ atoms within 12Å, excluding contacts between $C_\alpha$ atoms belonging to the same helix. We define a nest as a region in which the collection of sidechain atoms of the partner $C_\alpha$ atoms can nestle. Panel a) shows two nests formed by the poking contacts of residue VAL-13 and LEU-40. In both cases, the number of $C_\alpha$ atoms involved in the nest is 7. The number

of relevant poking contacts are 11 and 15 for VAL-13 and LEU-40 respectively. Panel b) shows 17 nests of size 7 and 8 (other nests are smaller), altogether comprising 46 of 149 amino acids in the myoglobin protein chain. The 17 nests originate from poking contacts from 6 LEU, 2 ALA, 2 MET, 2 PHE, 2 VAL, 1 TRP, 1 ILE and 1 GLY residue and involve 190 poking contacts in all. LEU, ALA, VAL, ILE, GLY are aliphatic; TRP, PHE are aromatic; and MET contains sulfur. The side chain atoms of the nestling amino acids are predominantly hydrophobic carbon and sulfur atoms. The exceptions are several oxygen and nitrogen atoms which either poke out towards the water or are compensated by poking towards another oppositely charged entity.

Here we will *postulate* that the complex interactions mediated by sidechains can be approximately captured as the sum of emergent pairwise poking interactions between the backbone $C_\alpha$ atoms within a range of 12Å, which are not involved in the formation of the common building blocks of protein structures (Figure 8). We will choose the simplest option of merely counting the number of such poking pairwise interactions and assign each of them a happy energy of around a fifth of a handholding pair. The range of 12Å is determined as roughly twice the size of the largest sidechains. We have verified that our results are essentially independent of the precise value of this range.

Poking contacts also play a vital role in a context unrelated to proteins when one has a hard-core constraint on the closest approach of two segments of a chain. To ensure self-avoidance, one would look for all poking contacts and ensure that the pairwise distance in the closest one among these is no smaller than the hardcore constraint. There is a vexing problem pertaining to ensuring self-avoidance of a continuum chain, where local particles along the chain are necessarily very close to each other because of the tethering. Any pairwise potential with an energy penalty for too close an approach between particles would simply not work in the continuum limit because of the large number of close by contiguous contacts along the chain. This problem can be deftly solved by discarding a pairwise potential and instead employing a three-body potential [16].

Poking interactions provide an alternative way of addressing this problem. Starting from any point on a chain (continuum or discrete), the spatial distance when plotted against sequence separation can exhibit local minima at non-zero sequence separation, which signal poking interactions, as long as there is reciprocity. The geometry of such a graph can be used to separate out local and non-local contacts along a chain with all contacts within the first maximum being local. Such local contacts are a natural attribute of any chain and do not have to accounted for while dealing with self-avoidance.

There are two significant advantages of poking contacts in a chain context. First, they are entirely compatible with the interactions used for the construction of building blocks – there is no frustration$^2$ in the nature of the interaction. Second, in the building blocks, there were just two hands available for holding, thus permitting anisotropic structures like the one-dimensional helix and strand and the two-dimensional sheet. Even in the assembly process, the socialist nature of indiscriminate interactions with everyone interacting with everyone else who is close by, is replaced with a manageable number of emergent poking interactions. This simplifies the model and ties the relevant interactions directly to the geometry of the structure. The concept of poking interactions in the polymer field and their utility in protein science at all levels of assembly (secondary and tertiary structure) has not been considered before.

We end this section with a recapitulation of the key ideas. We show that the helix and the sheet both result from handholding of the coins associated with the backbone atoms. This imposes certain geometrical constraints on the sites containing the paired coins. In particular, the handholding in both building blocks is associated with poking pairwise interactions. The sidechains do not play a role in determining the structure of the common building blocks. We suggest that the assembly process, which is driven by sidechain interactions, can be captured effectively as a sum of pairwise poking contacts of the $C_\alpha$ atoms without worrying about sidechain specificity. We thus end up with a homopolymer model and we will explore its energy landscape further.

## 3.4 Computer model of a protein chain

Armed with these insights, we turn to the simulation of a chain with a uniform bond length, b, of around 3.81Å. b sets the length scale in the problem and is chosen to match that of proteins to allow us to make quantitative comparisons with protein data. We impose a constraint on the local bond-bending angle that it cannot be smaller than $\theta_{min} \sim 91°$. We also require that no pair of $C_\alpha$ atoms can be within a hard-core distance of around 4.5 Å (derived from the van der Waals radius of an isolated glycine amino acid [85]) from each other. We have verified that our results are substantially independent of these two choices.

The intra-helix contacts are pairwise poking contacts of the (i,i+3) type with the correct constraints to ensure Kepler touching of the coins. We impose soft constraints on the values of the dot products $\mathbf{t_i} \cdot \mathbf{b_{i+3}}$ and $\mathbf{b_i} \cdot \mathbf{t_{i+3}}$, as well as the dot products $\mathbf{t_i} \cdot \mathbf{r_{i,i+3}}$ and $\mathbf{t_{i+3}} \cdot \mathbf{r_{i,i+3}}$. Here $\mathbf{t_i}$ and $\mathbf{b_i}$ represent the tangent and binormal vectors at bead i in the local Frenet system of coordinates [80] and $\mathbf{r_{i,i+3}}$ is the vector connecting beads i and i+3. Specifically, we assign no α-basin reward unless the dot products $\mathbf{t_i} \cdot \mathbf{b_{i+3}}$ and $\mathbf{b_i} \cdot \mathbf{t_{i+3}}$, as well as $\mathbf{t_i} \cdot \mathbf{r_{i,i+3}}$ and $\mathbf{t_{i+3}} \cdot \mathbf{r_{i,i+3}}$, lie in appropriate ranges that are deduced from the *righthanded* Kepler helix. The dot products $\mathbf{t_i} \cdot \mathbf{b_{i+3}}$ and $\mathbf{b_i} \cdot \mathbf{t_{i+3}}$ are both required to lie between +0.156 and +0.325 allowing for ±5° tolerance around the ideal angle of ~76° between the corresponding vectors. Likewise, the dot products $\mathbf{t_i} \cdot \mathbf{r_{i,i+3}}$ and $\mathbf{t_{i+3}} \cdot \mathbf{r_{i,i+3}}$ need to lie in the range between +0.127 and +0.297, permitting a ±5° tolerance around the ideal angle of ~77.7° between the corresponding vectors in the Kepler helix.

The zigzagging of an individual strand is ensured by considering the relative orientations of $\mathbf{n_i}$ and $\mathbf{n_{i+1}}$, where $\mathbf{n_i}$ is the normal vector at site i [20]. Ideally, these vectors ought to be antiparallel, but in our simulations, we merely require that the angle between them is at least 120°. In addition, we require a nearly perpendicular orientation of the connecting vector between i and j (that are non-local poking contacts, with j>i+3), $\mathbf{r_{ij}}$, with both $\mathbf{n_i}$ and $\mathbf{n_j}$ to account for the geometries of the paired axes in a β-sheet. In our simulations, we allow for a tolerance of ±5° around the 90° angle for ideal strands. Our computer model is deliberately simplified for handholding in sheets. It corresponds to Figure 5a and leads to ideal sheets. There is no distinction between parallel and antiparallel strand pairing and there is no possibility of squeezing.

In summary, our computer simulation model does away with coins and is in the same spirit as that presented originally by Hoang et al. [20]. The common idea is to capture the features of protein native state structures, independent of amino acid sequence, through suitable Frenet constraints within the context of a tube model. In Ref. [20], the soft constraints were determined primarily by a detailed analysis of PDB data. Here again we use empirical data (most notably that naturally occurring helices are right-handed) but our constraints are derived from theory of the Kepler helix and sheet and are not directly based on protein data. Most importantly, our work is built on the observation that poking contacts play a critical role in handholding in helices and sheets along with our hypothesis that poking contacts may be important in the assembly of the secondary motifs as well. Our simulations are therefore strictly unrelated to proteins, they are based on geometry and yet turn out to yield results like proteins.

Operationally, we assign *all* poking contacts within 12Å with a reward of -$E_\gamma$. We consider two special cases, exclusive of each other, of poking contacts (i,j) that correspond to handholding within the secondary structures, which we additionally reward. These (i,j) poking contacts are predicted to have an ideal value of 2Δ=5.26Å. In our simulations, we allow for poking contacts to be within 6Å. The first case is relevant for helix handholding and involves (i,i+3) contacts with soft constraints on the values of the dot products, $\mathbf{t_i}\cdot\mathbf{b_{i+3}}$, $\mathbf{b_i}\cdot\mathbf{t_{i+3}}$ and $\mathbf{t_i}\cdot\mathbf{r_{i,i+3}}$ and $\mathbf{t_{i+3}}\cdot\mathbf{r_{i,i+3}}$. Such contacts are allocated an additional -($E_\alpha$ - $E_\gamma$) reward. The second case involves (i,j) contacts with j>i+3, with soft constraints on the dot products $\mathbf{n_i}\cdot\mathbf{n_{i-1}}$, $\mathbf{n_i}\cdot\mathbf{n_{i+1}}$, $\mathbf{n_j}\cdot\mathbf{n_{j-1}}$, $\mathbf{n_j}\cdot\mathbf{n_{j+1}}$, $\mathbf{n_i}\cdot\mathbf{r_{ij}}$, and $\mathbf{n_j}\cdot\mathbf{r_{ij}}$. These contacts are rewarded additionally by an amount -($E_\beta$ - $E_\gamma$).

Our *homopolymer* model has three kinds of favorable pairwise interactions: helix handholding (with an energy score of -$E_\alpha$ = -1 for each such contact), sheet handholding (with an energy score of -$E_\beta$ = -1 per contact) and poking pairwise nestling interactions within 12Å that are not part of a helix or a sheet (with an energy score of -$E_\gamma$ = -0.2 per pairwise contact). We explore the conformation space of this model and determine those conformations with overall large happiness or low energy score. Except for boundary effects, the fully satisfied helix and sheet are degenerate because each backbone $C_\alpha$ atom has two hands. Assemblies of one-dimensional helices and two-dimensional sheets accrue favorable nestling contacts leading to protein-like modular assemblies of helices and sheets connected by loops. There is a significant number of degenerate low energy conformations, and we show four of these in Figure 9. We do not pre-assign any given $C_\alpha$ atom to a helix, a sheet, or to a nestling contact. A given atom selects the most favorable accessible assignment, depending on its environment, spontaneously yielding the high degeneracy. All low energy structures are protein-like modular structures because of the energy balance between the more favorable helix and sheet contacts and less competitive nestling contacts. Heterogeneity in the chain will favor the best-fit structure for the sequence over the other structures with relative ease.

Figure 9 shows a few representative protein-like structures with low-lying energies arising from our simulations. We obtain nearly degenerate modular structures made up of building blocks of helices and strands assembled into sheets. It is important to note that our simulations were carried out for individual homopolymers, of modest size 80, with no side chains. At moderate temperatures, we observe frequent spontaneous switching between multiple low energy protein-like conformations, including ones not shown here. The energy landscape is studded with numerous minima corresponding to building blocks assembled with different topologies. That this happens at the homopolymer level is at odds with the common belief that sequence determines structure. Instead, as suggested earlier [20], the sequence does not have the onerous task of sculpting a folding funnel [13-15] from scratch but rather merely needs to refine the presculpted landscape to enhance the fit of the sequence into the best choice native state within the existing library of folds. The menu of putative protein-like native structures has been computationally explored in homopolymer chains employing atomistic force fields [86-88] and density functional-based mean-field potentials [89] and underscore the secondary role of the sequence in sculpting the fold.

## 4. Conclusions

The protein problem is inherently complex with thousands of atoms tethered to each other surrounded by solvent molecules. There are twenty types of amino acids. There are numerous types of interactions. And there is the ever-present role of evolution. Yet, within this complexity, there is simplicity [20] provided by the topologically limited number of protein folds determined by physical law, geometry, and symmetry, in the backdrop of which sequences and functionalities are shaped by evolution.

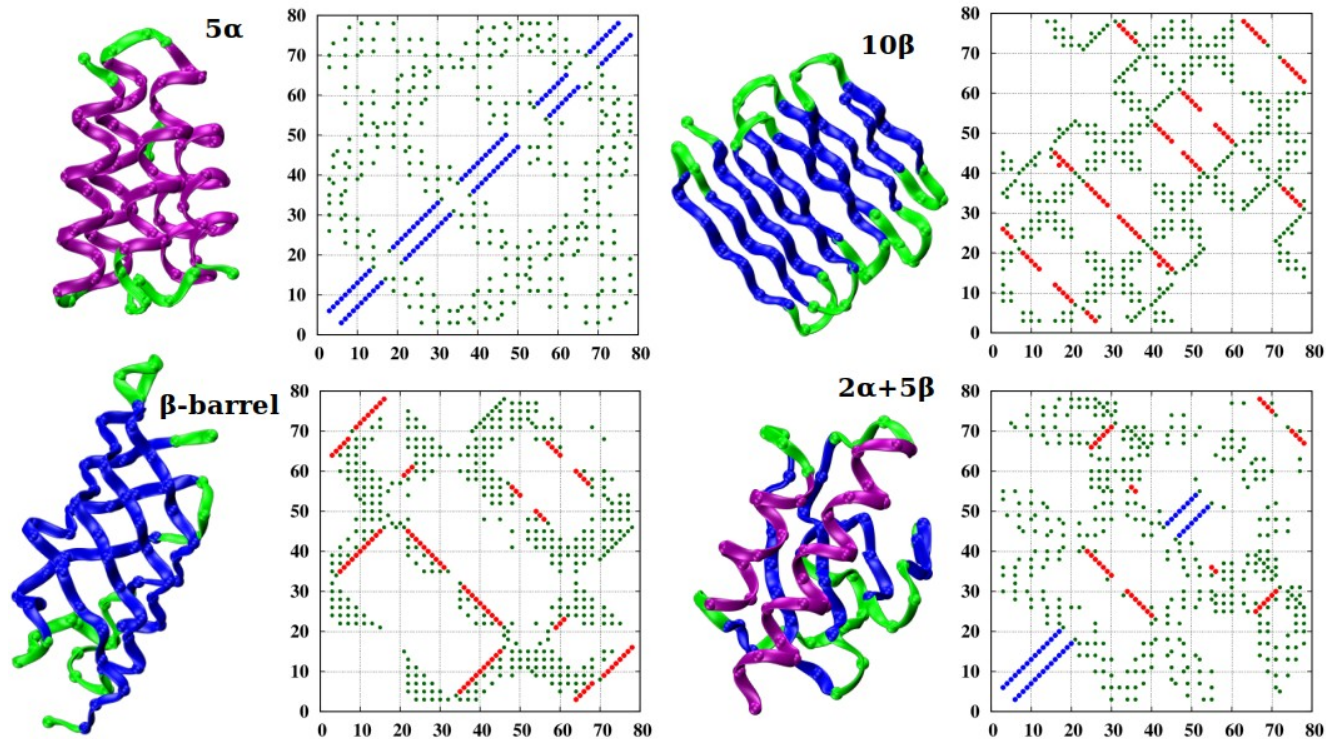

Figure 9: Four low energy structures with distinct topologies. We used parallel tempering to determine low energy conformations for a homopolymer chain of length 80. Starting from the upper left and proceeding in a clockwise direction, we show a five-helix bundle (5α), a β-sheet structure (10β) comprised of two approximately parallel planes, each having 5 β-strands, a (2α+5β) structure with two α-helices lying on top of a β-sheet comprising five strands, and finally a β-barrel structure, comprised of β-strands arranged in a cylindrical fashion. Helices are shown in purple, β-strands in blue, and loops in light green. Structures are drawn in ribbon representation and positions of $C_\alpha$ atoms are shown as spheres. We show the corresponding contact maps alongside the structures. The color code employed in contact maps is blue for the Kepler helix backbone-backbone poking contacts, red for the Kepler β backbone-backbone poking contacts, and dark green for the nestling poking pairwise

contacts within 12Å. For clarity, we have shrunk the sizes of points representing the nestling contacts. The energies of the four structures are approximately equal in magnitude (-84, -82, -86, and -81) for $E_α = 1$, $E_β = 1$, and $E_γ = 0.2$.

Here we study the geometry of native state structures of a chain. Our analysis applies only to proteins because our assumptions are specialized to this context. We use the fact that all proteins share the same backbone. The backbone interactions yield the well-known building blocks shared by all proteins. We suggest that the assembly of building blocks may be captured by means of poking pairwise contacts. We find that the geometries of the building blocks and the tertiary structure of our simplified model (Figure 9) are in good accord with those of real proteins.

Our analysis is in the same vein as Kepler's study of the packing of cannonballs in the hold of a ship (or the equivalent problem of how a grocer ought to pack her apples) [67-69]. Kepler correctly conjectured that no structure was better than the fcc structure for the packing *and* space-filling of spheres. A sphere is isotropic and changing the symmetry to study the packing of cubes instead results in an optimal simple cubic lattice structure. This underscores the key role played by symmetry in space-filling. Idealized crystals are space-filling, infinite in extent, and periodic. What works at one location works at another leading to translational invariance.

Our analysis here corresponds to a study of the Kepler-like conformations of a finite sized chain, accounting for symmetry and geometry information whenever available, and can explain the common characteristics of all globular proteins. Thus, in a very real sense, the protein structure problem has the same ingredients as the centuries-old problem of a grocer arranging her apples.

We conclude by celebrating the uniqueness of proteins. First, despite their modest sizes, proteins exhibit many common characteristics. This is distinct from universality in critical phenomena [90] where, in the long length scale limit, many details become irrelevant and power law behavior has universal exponents. Thus, when the dimensionality and the symmetry of ordering are the same, many details are irrelevant and disparate systems such as a liquid-vapor critical point, a binary alloy at the onset of ordering, and a spin system with up-down symmetry on a three-dimensional lattice at its ordering temperature all exhibit the same critical behavior. In contrast, the common characteristics of proteins arises because all proteins have the same backbone and handholding considerations at the backbone level. This leads to the same building blocks for all proteins and to modular native state folds.

Proteins exhibit stability and diversity. A classic physics example of such behavior is the spin glass phase [91,92] where frustration (the inability to satisfy all interactions simultaneously) results in a rugged energy landscape characterized by many local minima [93]. The diversity of low energy states along with their individual stability has found direct use in models of prebiotic evolution [94,95] and content addressable memories [96]. Proteins also exhibit stability and diversity in the presculpted landscape of a homopolymer. These attributes do not originate from frustration and a protein is not necessarily plagued by sluggish dynamics associated with being trapped in a hugely rugged landscape [97-99]. Unlike a spin model (which is what a spin glass is), protein structures are actual three-dimensional sculptures, which directly host and facilitate interactions within the living cell.

Like liquid crystals [67,100], proteins exhibit sensitivity. Liquid crystals are sometimes referred to as the most sensitive phase of matter. Enzymes are incredibly versatile and can speed up reactions by many orders of magnitude. Unlike a liquid crystal built up of anisotropic constituents like rods [see for example, 101] or banana shaped molecules, here the anisotropy is inherent because of the tethering along a chain. Furthermore, the building blocks themselves are anisotropic because of handholding, which is anisotropic. A chain provides a powerful context of where an object is along it, not available in a liquid crystal. Just as a liquid crystal derives its sensitivity by being poised in the vicinity of a phase transition to the liquid state, here the space filling conformations of coins in this finite sized system are automatically poised in a marginally compact phase in the vicinity of a swollen phase. The marginally compact phase is characterized by the perfect match between the tube size (the coin diameter) and the interaction range for two coins just touching each other (also the coin diameter). The sensitivity of proteins is underscored even in our rudimentary simulations where a homopolymer dynamically switches from structure to structure due to thermal fluctuations. Our work opens the possibility of the creation of functional entities at the nanoscale, based on ideas from geometry and symmetry, which can switch reversibly between distinct geometries and exhibit novel emergent behavior, when networked together.

Pauling and his colleagues [5] used "*the complete and accurate determination of the crystal structure of amino acids, peptides, and other simple substances related to proteins …. to construct two reasonable hydrogen-bonded helical configurations for the polypeptide chain*". In an unrelated vein, Ramachandran and his colleagues showed that the need to avoid steric overlaps of the backbone atoms permit and promote the existence of helices and sheets [7,8]. Here we have used geometry and symmetry to show that helices and sheets can also be viewed as arrangements of touching coins. Is this evidence for fine-tuning in nature, which permits distinct approaches to converge to the same results? Furthermore, the tube diameter, determined by the details of the backbone atomic structure, is required to be fine-tuned to a value of around 5.26Å to facilitate the Keplerian building blocks. The modularity of protein structures is a direct consequence of the backbones of proteins shaping the building blocks. The virtually perfect fit of quantum chemistry, e.g., the planarity of the peptide bond, the lengths of the covalent and hydrogen bonds, hydrophobicity, and steric constraints to be compatible with and promote protein native state structures is striking. That geometry and symmetry considerations yield the same structures is astonishing.

Proteins are unique in exhibiting stability, diversity, and sensitivity with geometrically well-defined native state structures. Alas, they are also able to ag-gregate and clump together as an insoluble amyloid. The quantum chemistry approach for studying proteins, which is surely correct albeit complex, and the simple Keplerian approach outlined here, which is a model and is therefore necessarily wrong, seem to be compatible with each other. We hope that our model will prove to be useful because of its simplicity. In future papers, we will present details and comparisons of our predictions with data on globular proteins, we will develop a simple picture of amyloid formation validated by experimental data, and we will elucidate the critical role played by side chains. An exciting recent development is the role of biomolecular condensates to create membrane-less compartments within a cell through liquid-liquid phase separation, facilitated by protein-protein and/or protein-RNA interactions [102]. It is an intriguing possibility that poking interactions may also play a role in this arena as well.

## Author contributions

## Acknowledgements

We are indebted to George Rose for his inspiration and fruitful collaboration. One of us (JRB) learned about proteins during multiple visits to GNR's Molecular Biophysics Unit at the Indian Institute of Science. We acknowledge joyous collaborations with Saraswathi Vishveshwara and helpful discussions with P. Balaram. The computer calculations were performed on the Talapas cluster at the University of Oregon.

## Funding information

This project received funding from the European Union's Horizon 2020 research and innovation program under Marie Skłodowska-Curie Grant Agree-ment No. 894784 (TŠ). The contents reflect only the authors' view and not the views of the European Commission. JRB was supported by a Knight Chair at the University of Oregon. TXH is supported by The Vietnam Academy of Science and Technology under grant No. NVCC05.05/22-23.

**Conflict of interest**

The authors declare that there is no conflict of interest.

**Endnotes:**

[1] In earlier work [36], we had derived the characteristics of a discrete space-filling helix with $\eta$ fixed and equal to that of the corresponding continuum helix and finding the best fit $\varepsilon_0$ for fulfilling the same conditions imposed here. These conditions were derived, not directly, as here, from consideration of the touching of coins but from the constraints inherent in the continuum helix. Remarkably, the results reported earlier are in excellent accord with those determined here correctly with the actual $\eta$ value differing from the continuum counterpart by less than a percent.

[2] Our hypothesis is an application of a generalized principle of minimal frustration [13] in a new context. The standard application of the principle points out that protein sequence design must be carried out thoughtfully to avoid frustrating tendencies in the amino acid interactions. Here, we choose a simple pairwise interaction potential that is entirely compatible with the formation of Keplerian helices and sheets with no frustrating tendency. A generic indiscriminate attraction would violate the principle by destabilizing both a helix and a strand.